\documentclass{moriond}

\newcommand{\Mpc}{{\rm Mpc}}

\newcommand{\ion}[2]{{\text{{\sc #1}\,{\sc #2}}}}

\newcommand{\nS}{n_{\rm S}}

\newcommand{\beq}{\begin{equation}}   %

\newcommand{\eeq}{\end{equation}}   %

\newcommand{\beqa}{\begin{eqnarray}}   %

\newcommand{\eeqa}{\end{eqnarray}}   %

\newcommand{\beal}{\begin{align}}

\newcommand{\enal}{\end{align}}

\newcommand{\bspl}{\begin{split}}

\newcommand{\espl}{\end{split}}

\newcommand{\bsub}{\begin{subequations}}

\newcommand{\esub}{\end{subequations}}

\newcommand{\bmulti}{\begin{multline}}   %

\newcommand{\beqm}{\begin{mathletters}}   %

\newcommand{\eeqm}{\end{mathletters}}   %

\newcommand{\Te}{T_{\rm e}}

\newcommand{\pot}[2]{#1 \times 10^{#2}}

\usepackage{amsmath}
\usepackage{txfonts}
\usepackage[numbers, compress, sort]{natbib}
\usepackage{hyperref}
\usepackage{wrapfig}

\def\be{\begin{equation}}
\def\ee{\end{equation}}
\def\bea{\begin{eqnarray}}
\def\eea{\end{eqnarray}}

\newcommand{\Photo}{}

\begin{document}
\title{\Large Science with CMB spectral distortions}

\author{J. Chluba}

\address{\scriptsize Department of Physics and Astronomy, Johns Hopkins University, Bloomberg Center 435, \\
3400 N. Charles St., Baltimore, MD 21218, USA}

\maketitle\abstracts{
The measurements of COBE/FIRAS have shown that the CMB spectrum is extremely close to a perfect blackbody. There are, however, a number of processes in the early Universe that should create spectral distortions at a level which is within reach of present day technology. In this talk, I will give a brief overview of recent theoretical and experimental developments, explaining why future measurements of the CMB spectrum will open up an unexplored window to early-universe and particle physics with possible non-standard surprises but also several guaranteed signals awaiting us.}

\section{Introduction}
Cosmology is now a precise scientific discipline, with detailed theoretical models that fit a wealth of very accurate measurements. Of the many cosmological data sets, the cosmic microwave background (CMB) temperature and polarization {\it anisotropies} provide the most stringent and robust constraints to theoretical models, allowing us to determine the key parameters of our Universe and address fundamental questions about inflation and early-universe physics \citep{Smoot1992, WMAP_params, Planck2013params}.

But the CMB holds another, complementary and independent piece of invaluable information: its {\it frequency spectrum}. Departures of the CMB frequency spectrum from a blackbody -- commonly referred to as {\it spectral distortions} -- encode information about the thermal history of the early Universe (from when it was a few month old until today).
Since the measurements with COBE/FIRAS, the average CMB spectrum is known to be extremely close to a perfect blackbody, with possible distortions being limited to $\Delta I_\nu/I_\nu \lesssim 10^{-5}-10^{-4}$ \citep{Mather1994, Fixsen1996}. 
This impressive measurement was awarded the Nobelprize in Physics 2006, and already rules out cosmologies with extended periods of significant energy release, disturbing the thermal equilibrium between matter and radiation in the early Universe.

Given that no average CMB distortion was found, {\it why is it still interesting to think about CMB spectral distortions}? First of all, there is a long list of processes that could lead to spectral distortions. These include: {\it reionization} and {\it structure formation} \citep{Sunyaev1972b, Hu1994pert, Cen1999, Miniati2000, Refregier2000, Oh2003, Zhang2004}; {\it decaying} or {\it annihilating particles} \citep{Hu1993b, McDonald2001, Chluba2013fore, Chluba2013PCA}; {\it dissipation of primordial density fluctuations} \citep[e.g.,][]{Sunyaev1970diss, Daly1991, Barrow1991, Hu1994, Chluba2012, Pajer2012, Ganc2012, Chluba2012inflaton, Chluba2013iso}; {\it cosmic strings} \citep{Ostriker1987, Tashiro2012, Tashiro2012b}; {\it primordial black holes} \citep{Carr2010, Pani2013}; {\it small-scale magnetic fields} \citep{Jedamzik2000, Kunze2014}; the {\it adiabatic cooling of matter} \citep{Chluba2005, Chluba2011therm}; {\it cosmological recombination} \citep[for overview see][]{Sunyaev2009}; and some new physics examples \citep{Moss2011, Lochan2012, Bull2013, Brax2013, Bruck2013, Caldwell2013}. Most importantly, many of these processes (e.g., reionization and cosmological recombination) are part of our standard cosmology and therefore should lead to guaranteed signals to search for.

The second reason for spectral distortion being interesting is due to technological advances. Although measurements of the CMB temperature and polarization anisotropies have improved significantly in terms of angular resolution and sensitivity since COBE/DMR, our knowledge of the CMB spectrum is still in a similar state as more than 20 years ago. This could change dramatically in the future with experimental concepts like PIXIE \citep{Kogut2011PIXIE} and PRISM \citep{PRISM2013WPII} (or its smaller M4 class version, COrE+) being presently discussed by the cosmology community. These types of experiments could possibly improve the limits of COBE/FIRAS by more than three orders of magnitude, providing a unique way to learn about processes that are otherwise hidden from us. At this stage, CMB spectral distortion measurements are furthermore only possible from space, so that in contrast to $B$-mode polarization science competition from the ground is largely excluded, making CMB spectral distortions a unique target for future CMB space missions \citep{Chluba2014Science}. This immense potential of spectral distortions was also recently recognized in the NASA 30-year Roadmap study, where improved characterization of the CMB spectrum was declared as one of the future targets \citep{Roadmap2014}.

\begin{figure} 
   \centering
   \includegraphics[width=0.81\columnwidth]{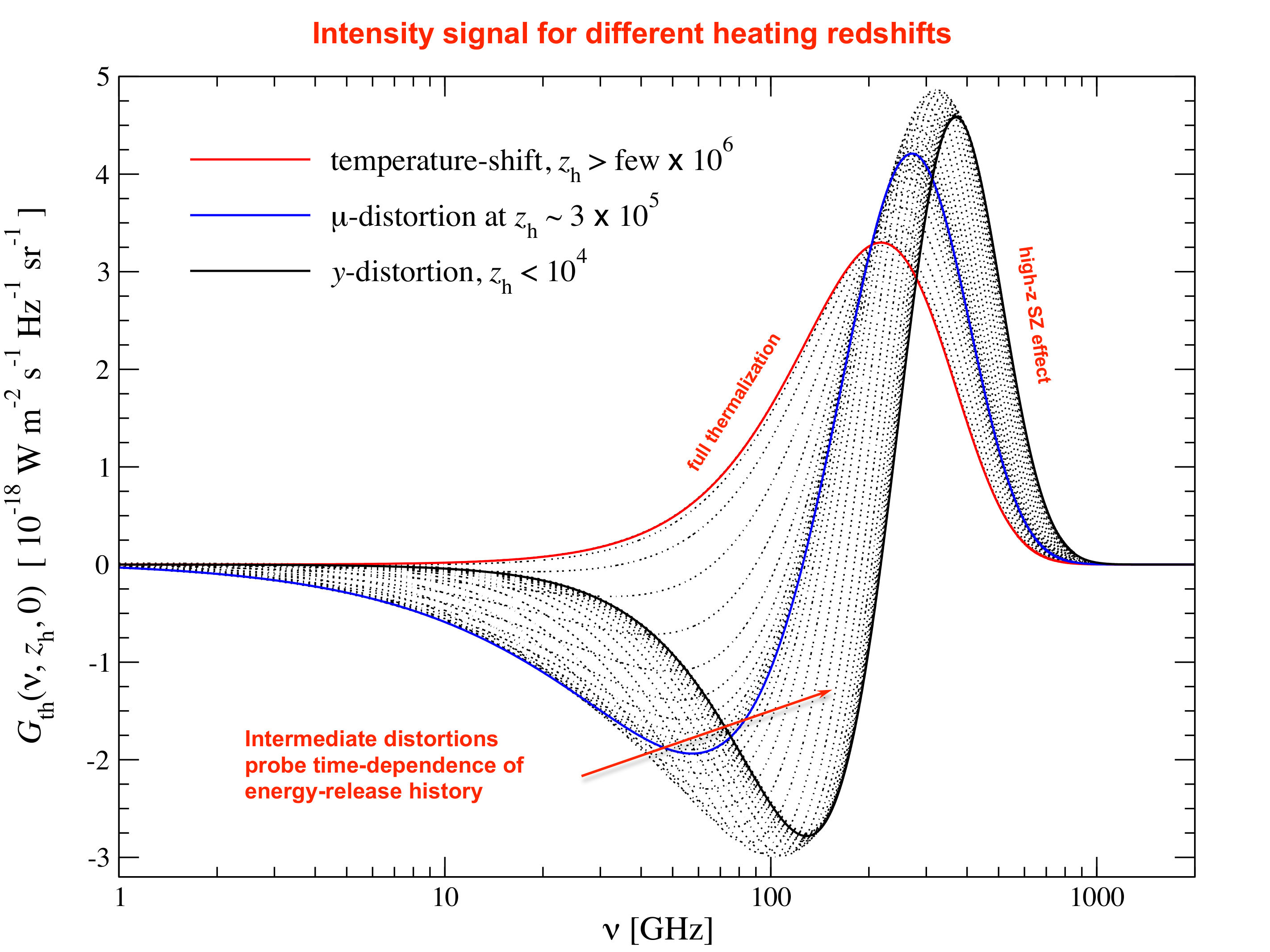}
   \caption{Change in the CMB spectrum after a single energy release at different heating redshifts, $z_{\rm h}$. At $z\gtrsim \pot{\rm few}{6}$, a temperature shift is created. Around $z \simeq \pot{3}{5}$ a pure $\mu$-distortion appears, while at $z\lesssim 10^4$ a pure $y$-distortion is formed. At all intermediate stages, the signal is given by a superposition of these extreme cases with a small residual (non-$\mu$/non-$y$) distortion that contains information about the time-dependence of the energy-release process (Figure adapted from \cite{Chluba2013Green}).}
   \label{fig:Greens}
\end{figure}
\vspace{-3mm}
\section{Thermalization physics and different types of primordial distortions}
It is well-known that energy release in the early Universe causes CMB spectral distortions \citep{Zeldovich1969, Sunyaev1970mu, Illarionov1975, Illarionov1975b, Danese1977, Burigana1991, Hu1993}. 
At redshifts $z\gtrsim \pot{\rm few}{6}$, the thermalization process, mediated by the combined action of double Compton emission, Bremsstrahlung and Compton scattering, is extremely rapid and erases any distortion to unobservable levels until the present, only leading to an increase of the average CMB temperature. The associated entropy production can be constrained using precise measurements of the light element abundances and the photon to baryon ratio \citep{Simha2008, Jeong2014Silk}.
At lower redshift, however, the CMB spectrum becomes vulnerable and a distortion remains. Traditionally this signal is described as a chemical potential $\mu$- and Compton $y$-distortion. 
A $\mu$-type distortion is created at very early epochs ($z\gtrsim \pot{5}{4}$) when redistribution of photons over frequency by Compton scattering with free electron is still very rapid, so that full kinetic equilibrium between electrons and photons can be achieved, producing a constant, non-vanishing chemical potential at high frequencies. A $y$-type distortion is produced in the other extreme, when energy exchange through Compton scattering is already inefficient and photons are only partially up-scattered, creating the high redshift ($z\lesssim \pot{5}{4}$) analogue of the thermal Sunyaev-Zeldovich (SZ) effect known from clusters of galaxies \cite{Zeldovich1969}.

\begin{figure} 
   \centering
   \includegraphics[width=0.78\columnwidth]{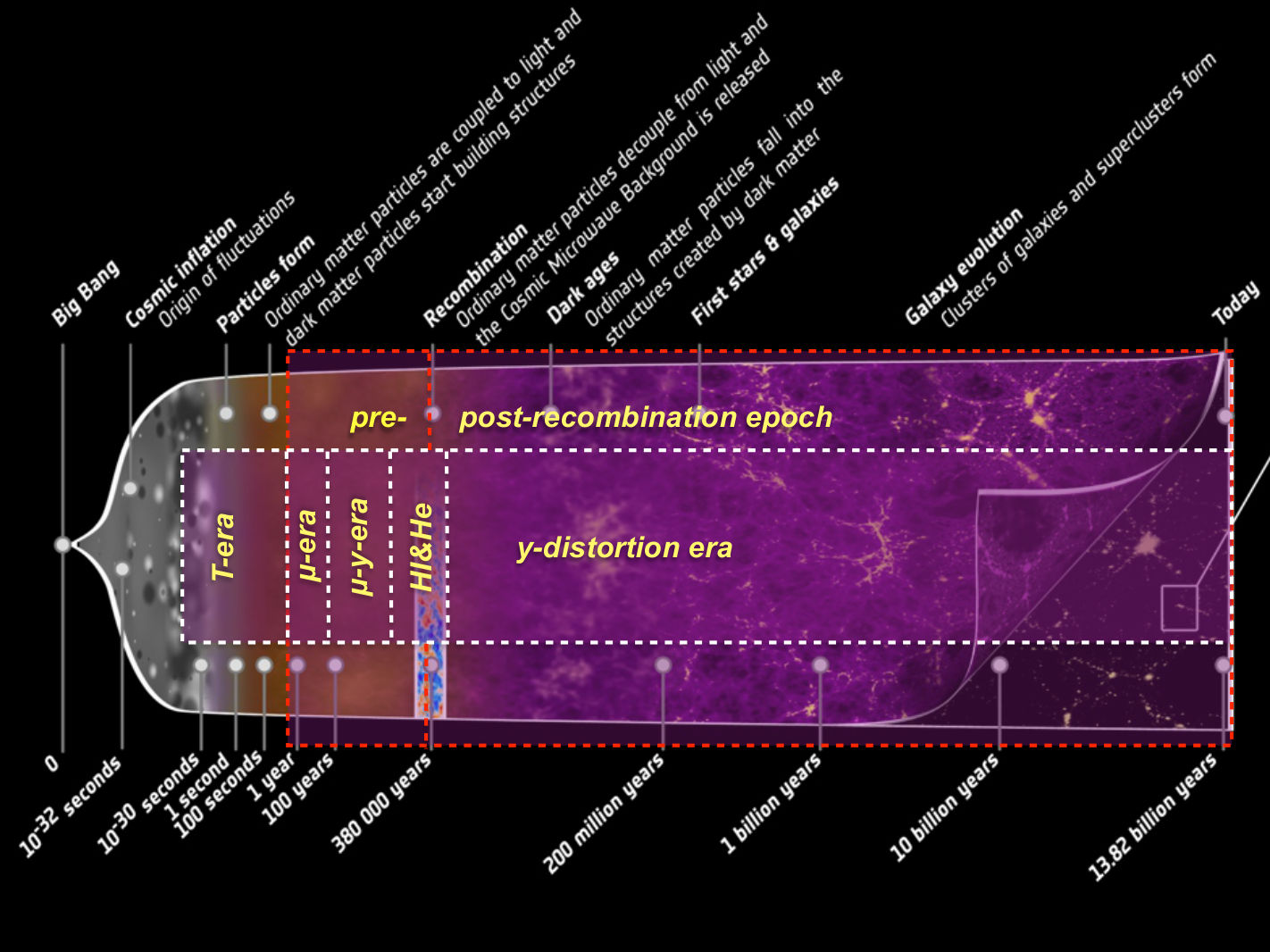}
   \caption{CMB spectral distortions probe the thermal history of the Universe at many stages during the pre- and post-recombination era. Energy release at $z\gtrsim \pot{\rm few}{6}$ only causes a change of the CMB temperature. A $\mu$-type distortion arises from energy release at $\pot{3}{5}\lesssim z\lesssim \pot{\rm few}{6}$, while a $y$-type distortions is created at $z\lesssim 10^4$. The signal caused during the $\mu/y$-transition era ($10^{4}\lesssim z\lesssim \pot{3}{5}$) is described by a superposition of $\mu$- and $y$-distortion with some small {\it residual} distortion that allows probing the time-dependence of the energy-release mechanism. In the recombination era ($10^3\lesssim z\lesssim 10^4$), additional spectral features appear due to atomic transitions of hydrogen and helium. These could allow us to distinguish pre- from post-recombination $y$-distortions (Figure adapted from \cite{PRISM2013WPII}).}
   \label{fig:history}
\end{figure}

It was, however, shown that the distortion signature from different energy-release scenarios is generally {\it not just} given by a superposition of pure $\mu$- and $y$-distortion \citep{Chluba2011therm, Khatri2012mix, Chluba2013Green}. The small residual beyond $\mu$- and $y$-distortion contains information about the {\it exact} time-dependence of the energy-release history (see Fig.~\ref{fig:Greens}), which in principle can be used to directly constrain, for instance, the shape of the small-scale power spectrum, measure the lifetime of decaying relic particles, or simply to discern between different energy-release mechanisms. In particular, \cite{Chluba2013fore, Chluba2013PCA} demonstrated that CMB spectrum measurement with a {\it PIXIE}-type experiment provide a sensitive probe for long-lived particles with lifetimes $t_{\rm X}\simeq 10^9\,{\rm sec}-10^{10}\,{\rm sec}$. Similarly, the shape of the small-scale power spectrum can be directly probed with {\it PIXIE}'s sensitivity if the amplitude of primordial curvature perturbations exceeds $A_\zeta\simeq \pot{\rm few}{-8}$ at wavenumber $k\simeq 45 \, \Mpc^{-1}$ \citep{Chluba2013fore, Chluba2013PCA}. 
CMB distortion measurements thus provide an unique way for studying early-universe models and particle physics at very different stages of the Universe (see Fig.~\ref{fig:history}).

These aspects are now rather well understood, and efficient methods for computing the CMB spectral distortions from any energy-release scenario exist \citep{Chluba2011therm, Chluba2013Green}. Information from the {\it residual} (non-$\mu$/non-$y$) distortion can be used to probe the time-dependence of processes occurring at $10^4\lesssim z \lesssim \pot{3}{5}$, going well beyond the less informative statement that energy was in fact liberated at some point \citep{Chluba2013PCA}.
However, the thermalization problem is even richer when including the effect of pre-recombination ($z\gtrsim 10^3$) atomic transitions \citep{Chluba2008c}. This might allow us to reach even deeper into the $\mu$- and $y$-eras by using spectral features of the cosmological recombination radiation \citep{Sunyaev2009}.

\section{CMB spectral distortion constraints for various scenarios}
In this part we briefly highlight some interesting scenarios that can be constrained using future CMB spectroscopy. We selected a few examples to illustrate the potential of CMB spectral distortions. For more in depth reading and overview we refer to \cite{Chluba2011therm}, \cite{Sunyaev2013}, \cite{Chluba2013fore} and \citep{Chluba2013PCA}.

\subsection{Reionization and structure formation}
Radiation from the first stars and galaxies \citep{Hu1994pert, Barkana2001}, feedback by supernovae \citep{Oh2003} and structure formation shocks \citep{Sunyaev1972b, Cen1999, Miniati2000} heat the IGM at redshifts $z\lesssim 10-20$, 
producing hot electrons (with temperatures $\Te\simeq 10^4\,{\rm K}-10^5\,{\rm K}$) that up-scatter CMB photons, giving rise to a Compton $y$-distortion with typical amplitude $\Delta I_\nu/I_\nu \simeq 10^{-7}-10^{-6}$. This signal will be detected at more than a $100\,\sigma$ with a PIXIE-type experiment, providing a sensitive probe of reionization physics and delivering a census of the missing baryons in the 
local Universe. A PRISM-like experiment may furthermore have the potential to separate the spatially varying 
signature caused by the WHIM and proto-clusters \citep{Refregier2000, Zhang2004}.

\begin{figure} 
   \centering
   \includegraphics[width=0.78\columnwidth]{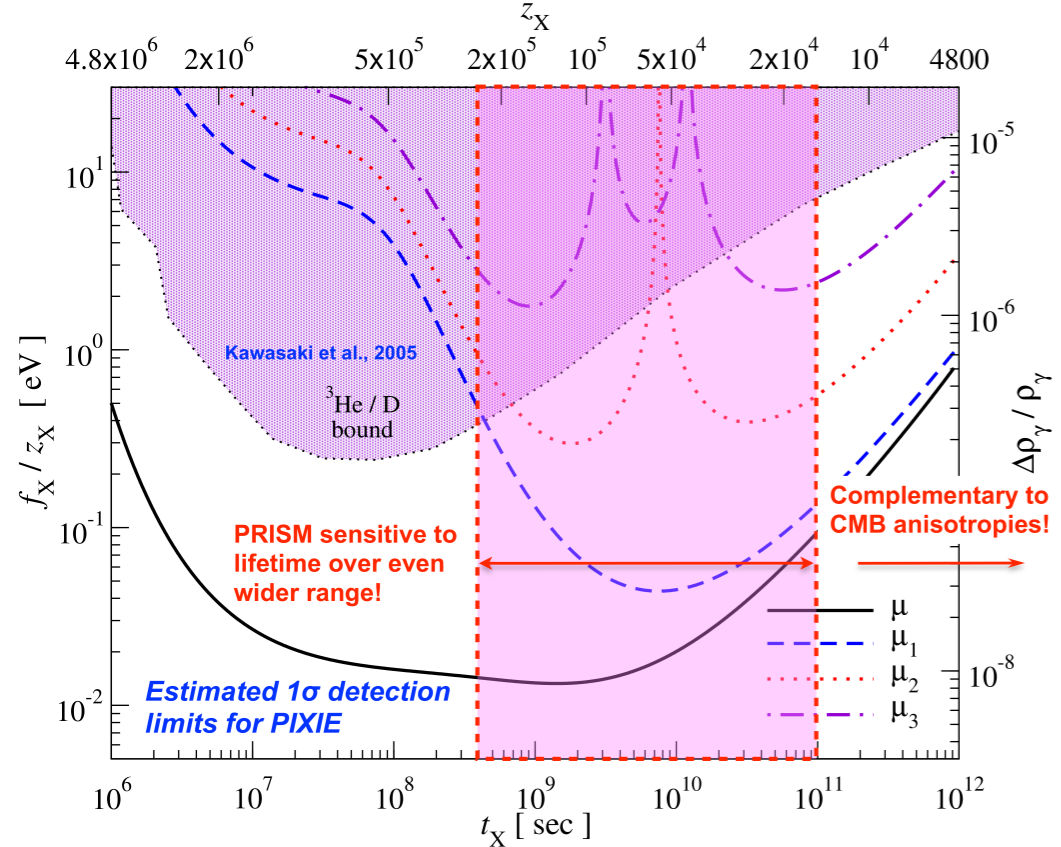}
   \caption{Decaying particle detection limits ($1\sigma$) for a PIXIE-like experiment. The eigenamplitudes $\mu_i$ characterize the non-$\mu$/non-$y$ distortion signal \citep{Chluba2013PCA}, which provides time-dependent information of the energy release history. CMB distortion limits could be $\simeq 50$ times tighter than those derived from light element abundances \citep{Kawasaki2005}. A separate determination of lifetime and particle abundance could be possible for lifetimes $t_{\rm X} \simeq 10^{8}\,{\rm sec}-10^{11}\,{\rm sec}$, being complementary to constraints derived using the CMB anisotropies \citep[e.g.,][]{Chen2004, Zhang2007}. The figure is adapted from \cite{Chluba2013PCA}.} 
   \label{fig:decay}
\end{figure}
\subsection{Decaying and annihilating particle scenarios}
The CMB spectrum allows us to place stringent limits on decaying and annihilating particles in the pre-recombination epoch \citep{Hu1993b, McDonald2001, Chluba2010a, Chluba2011therm}. This is especially interesting for decaying 
particles with lifetimes $t_{\rm X} \simeq 10^{8}\,{\rm sec}-10^{11}\,{\rm sec}$ \citep{Chluba2013fore, Chluba2013PCA}, as the exact shape of the distortion encodes when the decay occurred.
Decays or annihilations associated with significant low-energy photon production furthermore 
create a unique spectral signature that can be distinguished from simple energy release \citep{Hu1993}.
This would provide an unprecedented probe of early-universe particle physics (e.g., dark matter in excited states \citep{Pospelov2007, Finkbeiner2007} or Sommerfeld-enhanced annihilations close to resonance \citep{Hannestad2011}), with many natural particle candidates found in supersymmetric models \citep{Feng2003, Feng2010}.
For decaying particle scenarios, the $1\sigma$ detection limits expected for a PIXIE-like experiment are illustrated in Fig.~\ref{fig:decay}.

\begin{figure} 
   \centering
   \includegraphics[width=\columnwidth]{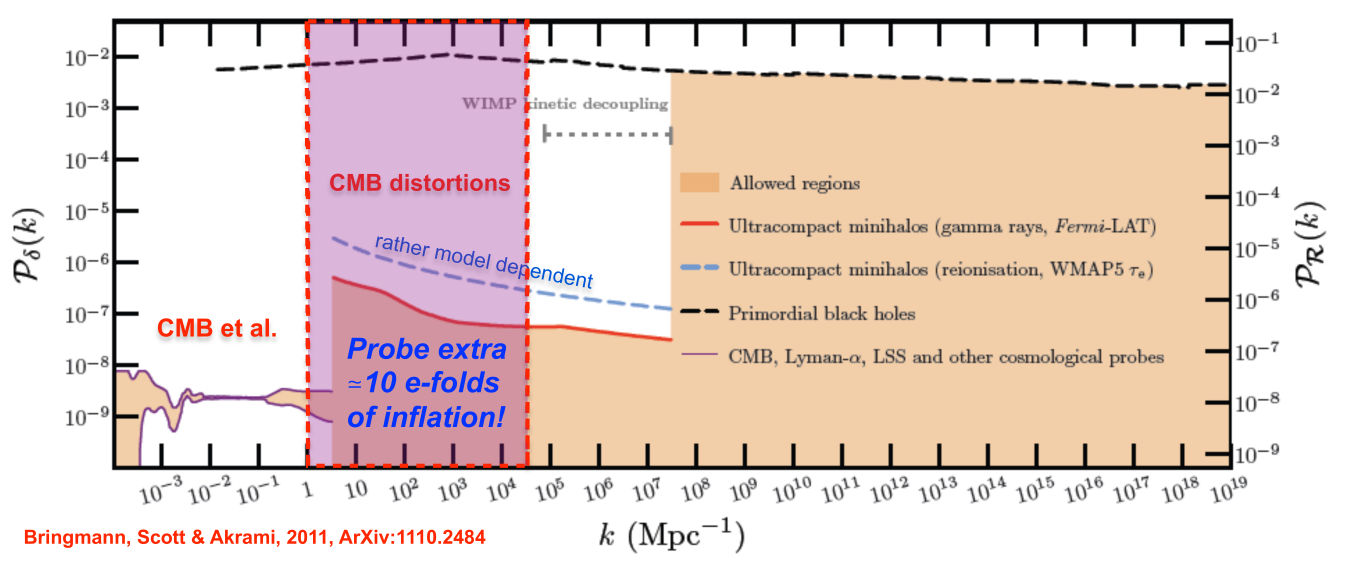}
   \caption{Current constraints on the small-scale power spectrum. At large scales ($k\lesssim 3\,{\rm Mpc}^{-1}$), CMB anisotropies and large scale structure measurements provide very stringent limits on the amplitude and shape of the primordial power spectrum. At smaller scales, the situation is more uncertain and at $3\,{\rm Mpc}^{-1}\lesssim k\lesssim 10^4\,{\rm Mpc}^{-1}$ which can be targeted with CMB spectral distortion measurements wiggle room of at least two orders of magnitude is present. CMB distortion measurements could improve these limits to a level similar to the large-scale constraints. The figure is adapted from \cite{BSA11}.} 
   \label{fig:constraints}
\end{figure}
\subsection{Dissipation of small-scale perturbations}
Silk-damping of small-scale perturbations in the photon fluid gives rise to CMB distortions \citep{Sunyaev1970diss,Daly1991,Barrow1991,Hu1994} which directly depend on 
the shape and amplitude of the primordial power spectrum at scales $0.6\,{\rm kpc}\lesssim 
\lambda \lesssim 1\,{\rm Mpc}$ (or multipoles $10^5\lesssim \ell \lesssim 10^8$) 
\citep{Chluba2012}. The physics of the mechanism is very simple and only related to the mixing of blackbodies with different temperatures by Thomson scattering, which initially creates a $y$-distortion \citep{Zeldovich1972, Chluba2004} that subsequently thermalizes.
This process allows constraining the trajectory of the inflaton at stages unexplored by ongoing or 
planned experiments \citep{Chluba2012inflaton, Powell2012, Khatri2013forecast, Chluba2013PCA}, extending our 
reach from 7 e-folds of inflation probed with the CMB anisotropies to a total of 17 e-folds. This is particularly interesting, because the experimental constraints on the small-scale power spectrum allow at least two orders of magnitude of wiggle room with respect to the constraints derived at large scales (wavenumber $k\lesssim 3\,{\rm Mpc}^{-1}$) from CMB and large scale structure measurements (see Fig.~\ref{fig:constraints}).

\begin{figure} 
   \centering
   \includegraphics[width=0.82\columnwidth]{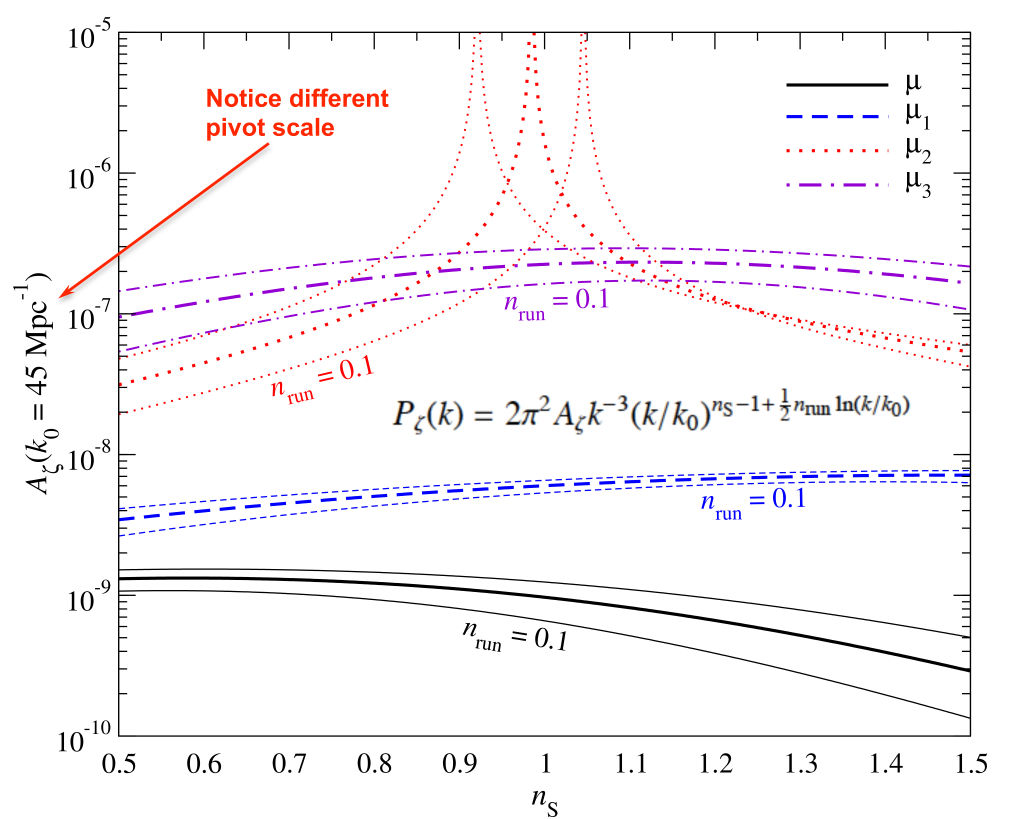}
   \caption{Detection limits ($1\sigma$) for a PIXIE-like experiment. CMB spectral distortion measurements could rule out early-universe models which create excess power at small scales above the level of $P_\zeta\simeq 10^{-9}$. A PRISM-like experiment may allow gaining an additional order of magnitude on this value. The figure is adapted from \cite{Chluba2013PCA}.} 
   \label{fig:dissipation}
\end{figure}

The distortion signal is also sensitive to the difference between adiabatic and isocurvature 
perturbations \citep{Barrow1991,Hu1994isocurv, Dent2012, Chluba2013iso}, as well as primordial 
non-Gaussianity in the ultra squeezed-limit, leading to a spatially varying spectral signal 
that correlates with CMB temperature anisotropies as large angular scales \citep{Pajer2012, Ganc2012}.
This effect therefore provides a unique way to study the scale-dependence of $f_{\rm NL}$ \citep{Biagetti2013}.
CMB spectral distortions hence deliver a complementary and independent probe of early-Universe physics, which allows capitalizing on the synergies with large-scale $B$-mode polarization measurements. The expected $1\sigma$ detection limits for a PIXIE-like experiment are illustrated in Fig.~\ref{fig:dissipation}.

\begin{figure} 
   \centering
   \includegraphics[width=0.82\columnwidth]{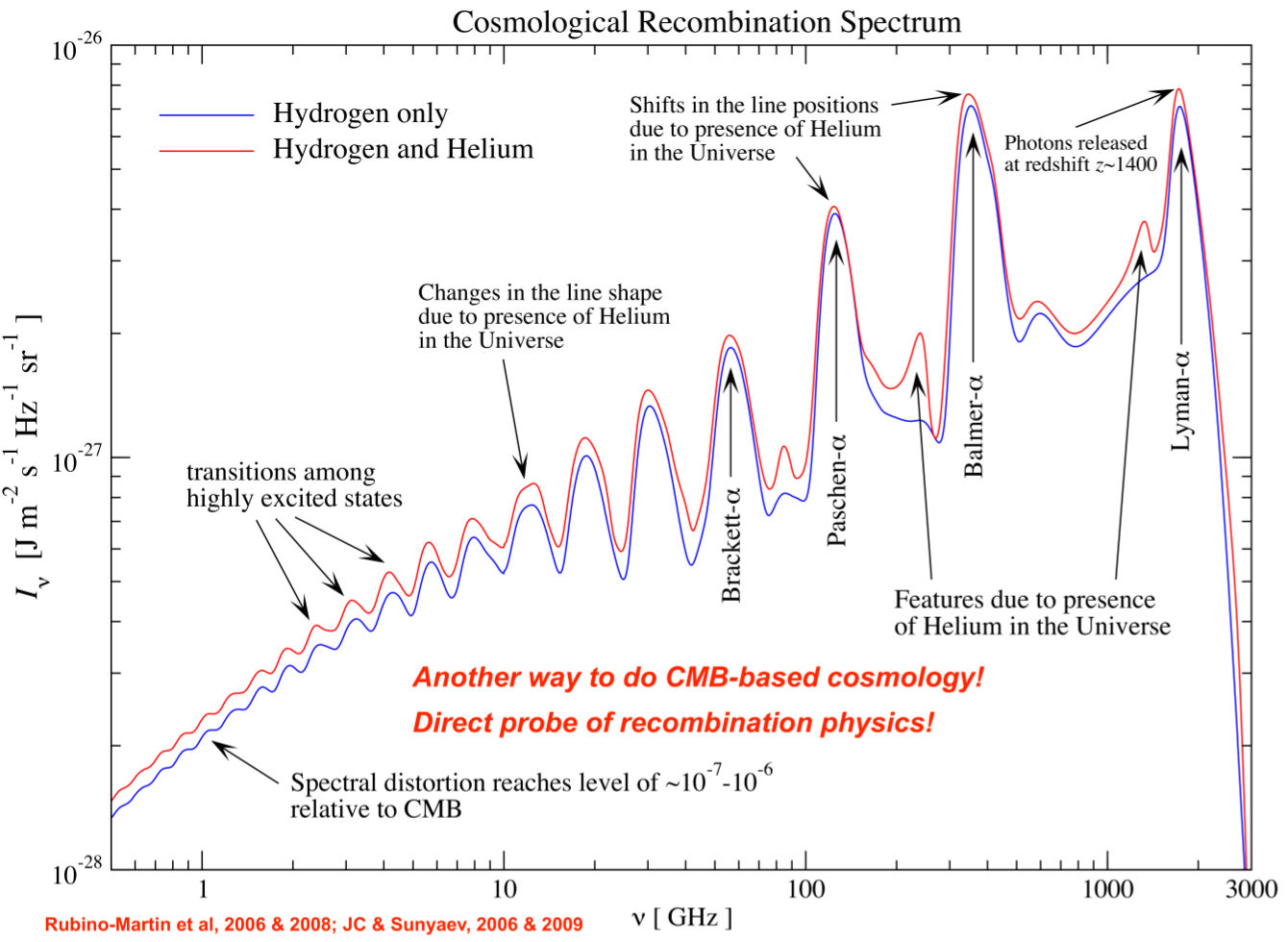}
   \caption{The cosmological recombination radiation created in the redshift range $z\simeq 10^3-10^4$. The presence of helium in the Universe gives rise to unique features in the recombination spectrum. This {\it fingerprint} of the recombination era in principle allows us to test our understanding of the recombination history which is one on the fundamental ingredients for the computations of the CMB anisotropies.} 
   \label{fig:recombination}
\end{figure}
\subsection{Cosmological recombination radiation}
The cosmological recombination of hydrogen and helium 
introduces distortions \citep{Zeldovich68, Peebles68, Dubrovich1975} at 
redshifts $z\simeq 10^3-10^4$, corresponding to $\simeq 260\,{\rm kyr}$ (\ion{H}{i}), $\simeq 
130\,{\rm kyr}$ (\ion{He}{i}), and $\simeq 18\,{\rm kyr}$ (\ion{He}{ii}) after the big bang 
\citep{Jose2006, Chluba2006b, Jose2008}. The overall 
signal is pretty small ($\Delta I_\nu/I_\nu \simeq 10^{-9}$) but its unique spectral features (see Fig.~\ref{fig:recombination}) promise an independent path to determination of cosmological parameters (like the baryon density and {\it pre-stellar} helium abundance) 
and direct measurements of the recombination dynamics, probing the Universe at stages before the last scattering surface \citep{Sunyaev2009}. Furthermore, if something unexpected happened during different stages of the recombination epoch, atomic species will react to this \citep{Chluba2008c} and produce additional distortion features that can exceed those of the normal recombination process. This will provide a unique way to distinguish pre- from post-recombination energy release \citep{Chluba2008c, Chluba2010a}.

To appreciate the importance of the cosmological recombination process at $z\simeq 10^3$, consider that today measurements of the CMB anisotropies are sensitive to uncertainties of the ionization history at a level of $\simeq 0.1\%-1\%$ \citep{Jose2010, Shaw2011}. For a precise interpretation of CMB data, uncertainties present in the original recombination calculations had to be reduced by including several previously omitted atomic physics and radiative transfer effects \citep[see][for overview]{Fendt2009, Jose2010}. This led to the development of the new recombination modules {\sc CosmoRec} \citep{Chluba2010b} and {\sc HyRec} \citep{Yacine2010c} which are used in the analysis of Planck data \citep{Planck2013params}. Without these improve treatments of the recombination calculation the value for $\nS$ would be biased by $\Delta \nS\simeq -0.01$ to $\nS\simeq 0.95$ instead of $\simeq 0.96$ \citep{Shaw2011}. We would be discussing different inflation models \citep{Planck2013iso} without these corrections taken into account! Conversely, this emphasizes how important it is to experimentally confirm the recombination process and CMB spectral distortions provide a way to do so. 

\section{Conclusions}
CMB spectral distortion measurements provide a unique way for studying processes in the pre- and post-recombination era. In the future, this could open up a new unexplored window to early-universe and particle physics, delivering independent and complementary pieces of information about the Universe we live in. 
We highlighted several processes that should lead to distortions at a level within reach of present-day technology. Different distortion signals can be computed precisely for various energy release scenarios.  Time-dependent information, beyond the standard $\mu$- and $y$-type parametrization, may allow us to independently constrain decaying particles and the shape and amplitude of the small-scale power spectrum of primordial perturbations. The cosmological recombination radiation will allow us to check our understanding of the recombination processes at redshifts of $z\simeq 10^3$. It furthermore should allow us to distinguish pre- from post-recombination $y$-distortions. 
This emphasizes the immense potential of CMB spectroscopy, both in terms of {\it discovery} and {\it characterization} science, and we should make use of this invaluable source of information with the next CMB space mission.  

\small
\bibliographystyle{mn2e}
\bibliography{Lit}

\end{document}